\documentclass[onecollarge,natbib]{svjour2}
\bibpunct{[}{]}{;}{n}{}{,} % to get "[numbered]" references from natbib
\smartqed  % flush right qed marks, e.g. at end of proof
\usepackage{graphicx}
\usepackage{amsmath}
\journalname{Few-Body Systems (EFB22)}
\begin{document}

\title{Exclusive $c \to s,d$ semileptonic decays of  spin-1/2 and spin-3/2  $cb$ baryons
%Insert your title here%\thanks{Grants or other notes
}


\author{C. Albertus         \and
        E. Hern\'andez       \and %etc.
        J. Nieves
}


\institute{C. Albertus \at
              Departamento de F\'\i sica At\'omica, Molecular y Nuclear. Facultad de Ciencias. Universidad de Granada.\\
              Av. de Fuentenueva S/N. E-18071, Granada, Spain\\
%              Tel.: +123-45-678910\\
%              Fax: +123-45-678910\\
              \email{albertus@ugr.es}           %  \\
           \and
           E. Hern\'andez \at
              Departamento de F\'\i sica Fundamental e IUFFyM. Universidad de Salamanca.\\
              Plaza de la Merced S/N. E-37008, Salamanca, Spain\\
              \email{gajatee@usal.es}
           \and
           J. Nieves \at
              Instituto de F\'\i sica Corpuscular, Centro Mixto CSIC-Universidad de Valencia\\
              Edificio de Institutos de Paterna.
              E-46071, Valencia, Spain\\
              \email{jmnieves@ific.uv.es}
}

\date{Received: date / Accepted: date}

\maketitle

\begin{abstract}
We present results for  exclusive semileptonic decay widths of
ground state spin-$1/2$ and spin-$3/2$ $cb$ baryons corresponding to a $c\to
s, d$ transition at the quark level. The relevance of hyperfine mixing in
spin-1/2 $cb$ baryons is shown. Our form factors are compatible with heavy quark
spin symmetry constraints obtained in the infinite heavy quark mass limit. 
\end{abstract}

\section{Introduction}
\label{intro}
This contribution summarizes the work of
Ref.~\cite{Albertus:2012jt}. There a systematic study
of exclusive semileptonic decay widths of
ground state spin-$1/2$ and spin-$3/2$ $cb$ baryons ,driven by a $c\to
s, d$ transition at the quark level, was done. Previous 
works~
\cite{SanchisLozano:1994vh,Faessler:2001mr,Kiselev:2001fw}
were limited to  just a few decay channels.

The baryons considered in this work are summarized in Table~\ref{tab:baryons}.
 The quark masses and wave functions have been calculated
with the AL1 potential of Ref.~\cite{Semay:1994ht}, using the 
variational
procedure described in \cite{Albertus:2006wb}.
\begin{table}
\begin{center}
\caption{Quantum numbers of baryons involved in this study. For the $cb$
baryons,
 states with a well defined spin for the heavy subsystem are shown.  $J^\pi$ and $I$ are the spin-parity and
 isospin of the baryon, while $S^\pi$ is the spin-parity of the two
 heavy or the two light quark subsystem. $n$ denotes a $u$ or $d$
 quark. Experimental masses are isospin averaged over the values reported by the
 Particle Data Group~\cite{Nakamura:2010zzi}.}
\label{tab:baryons}\begin{tabular}{ccccccc}\hline
Baryon &~~~~$J^P$~~~~&~~~~ $I$~~~~&~~~~$S^\pi$~~~~& 
Quark content &\multicolumn{2}{c}{Mass\ [MeV]}\\\cline{6-7}

       &       &         &   &      & Quark model   & Experiment                
\\    
\hline
$\Xi_{cb}$ &$\frac12^+$& $\frac12$ &$1^+$&$cbn$&6928&--
\\
$\Xi'_{cb}$ &$\frac12^+$& $\frac12$ &$0^+$&$cbn$&6958&--
\\
$\Xi^*_{cb}$ &$\frac32^+$& $\frac12$ &$1^+$&$cbn$&6996&--
\\
$\Omega_{cb}$  &$\frac12^+$& 0 &$1^+$&$cbs$&7013&--\\
$\Omega'_{cb}$  &$\frac12^+$& 0 &$0^+$&$cbs$&7038&--\\
$\Omega^*_{cb}$  &$\frac32^+$& 0 &$1^+$&$cbs$&7075&--\\

\hline
$\Lambda_b$ &$\frac12^+$& 0 &$0^+$&$udb$&5643&$5620.2\pm1.6$
\\
$\Sigma_b$ &$\frac12^+$& 1 &$1^+$&$nnb$&5851&$5811.5\pm2.4$
\\
$\Sigma^*_b$ &$\frac32^+$& 1 &$1^+$&$nnb$&5882&$5832.7\pm3.1$
\\
$\Xi_b$  &$\frac12^+$&$\frac12$&$0^+$&$nsb$&5808&$5790.5\pm2.7$
\\
$\Xi'_b$  &$\frac12^+$&$\frac12$&$1^+$&$nsb$&5946&--
\\
$\Xi^*_b$ &$\frac32^+$&$\frac12$&$1^+$&$nsb$&5975&--
\\
$\Omega_b$  &$\frac12^+$& 0 &$1^+$&$ssc$&6033&$6071\pm40$
\\
$\Omega^*_b$  &$\frac32^+$& 0 &$1^+$&$ssc$&6063&--
\\\hline
\end{tabular}
\end{center}
\end{table}
 In that table the double heavy baryon states have
been classified according to the total spin $S$ of the heavy quark subsystem. 
This is based in heavy quark spin symmetry (HQSS) that tell us that for very
large heavy quark masses one can select the heavy quark subsystem to have a well
defined total spin. However, and because of the finite value of the heavy 
quark masses, one has that for spin-1/2 $cb$ baryons, the hyperfine
interaction between the light quark and any of the heavy quarks can
admix both $S=0$ and $S=1$ components into their wave
function. The actual spin-1/2 physical states are
admixtures of the states $\Xi_{cb}$ and $\Xi_{cb}'$ ($\Omega_{cb}$ and
$\Omega_{cb}'$) given in Table~\ref{tab:baryons}. In Ref.~\cite{Albertus:2009ww}
we study this mixing finding the physical states and masses to be
\begin{align}
\Xi_{cb}^{(1)}&=0.431\Xi_{cb}-0.902\Xi'_{cb},
\ \  M_{\Xi_{cb}^{(1)}}=6967\, {\rm MeV}\ \ ;\ \ 
\Xi_{cb}^{(2)}=0.902\Xi_{cb}+0.431\Xi'_{cb},\ \  
M_{\Xi_{cb}^{(2)}}=6919\, {\rm MeV}\nonumber\\
\Omega_{cb}^{(1)}&=0.437\Omega_{cb}-0.899\Omega'_{cb},\ \ 
 M_{\Omega_{cb}^{(1)}}=7046\, {\rm MeV}\ \ ;\ \ 
\Omega_{cb}^{(2)}=0.899\Omega_{cb}+0.437\Omega'_{cb},\ \  M_{\Omega_{cb}^{(2)}}=7005\, {\rm MeV}
\end{align}
Mixing does not have a great impact on the masses, but the admixture
coefficients are large and mixing turns out to be very important 
for decay widths~
\cite{Roberts:2008wq,Albertus:2009ww,Albertus:2010hi}. It is worth noting that physical states are very
close to the states ($B$ stands for $\Xi$ or $\Omega$ in what follows)
\begin{eqnarray}
\widehat B_{cb}=-\frac{\sqrt3}{2}B'_{cb}+\frac{1}{2}B_{cb}\ \ ,\ \
\widehat B'_{cb}=\frac{1}{2}B'_{cb}+\frac{\sqrt3}{2}B_{cb}
\label{eq:qchqss}
\end{eqnarray}
in which it is the charm--light quark subsystem that has well defined
total spin $S_{cq}=1$ ($\widehat B_{cb}$) or $S_{cq}=0$ ($\widehat B'_{cb}$).
\section{Semileptonic decay widths}
Expressions for the decay widths and the form factor decompositions of the
hadronic matix elements can be found in Ref.~\cite{Albertus:2012jt}. For 
spin-$1/2$ to spin-$1/2$ transistions there are three vector ($F_1$, $F_2$,
  $F_3$) and three axial ($G_1$, $G_2$, $G_3$) form factors. For the case
  of spin-$1/2$ to spin-$3/2$ or spin-$3/2$ to spin-$1/2$ we have 
   four vector ($C_3^V$, $C_4^V$, $C_5^V$, $C_6^V$) and four axial
  ($C_3^A$, $C_4^A$, $C_5^A$, $C_6^A$) form factors. For spin-$3/2$ to 
  spin-$3/2$ a form factor
  decomposition can be found in~Ref.~\cite{Faessler:2009xn}. However in this
  latter case we do not calculate the form factor themselves but just the
  vector and axial matrix elements. 

Tables \ref{tab:resctos} and \ref{tab:resctod} summarize our results
for the semileptonic decay widths for $c\to s$ and $c\to d$
transitions. Results in brackets correspond to the case where
configuration mixing is ignored. As we see from the tables, in most of the 
cases,
configuration mixing greatly affects the decay widths. We also see our results
 agree with the few existing  previous calculations.
\begin{table}[h!!!]
\caption{Decay widths for $c\to s$ transitions. Results where
  configuration mixing is not considered are shown in parentheses. The
  result with a $\dagger$ corresponds to the decay of the $\widehat
  \Xi_{cb}$ state. The result with an ${\ast}$ is our estimate from
  the total decay width and the branching ratio given
  in~\cite{Kiselev:2001fw}. We have used $|V_{cs}|=0.97345$.
%Similar results are obtained for decays into
%$\mu^+\nu_\mu$.
} 
\label{tab:resctos}
\begin{center}
\begin{tabular}{llccc}\scriptsize
&\multicolumn{4}{c}{$\Gamma \ [10^{-14}\,{\rm GeV}]$}\\
&{This work}&\cite{SanchisLozano:1994vh}&\cite{Faessler:2001mr}&\cite{Kiselev:2001fw}\\
\hline
$\Xi^{(1)\,+}_{cbu}\to\Xi^0_b\, e^+\nu_e$& 3.74 (3.45)&(3.4)\\
$\Xi^{(2)\,+}_{cbu}\to\Xi^0_b\, e^+\nu_e$& 2.65 (2.87)\\
$\Xi^{(1)\,+}_{cbu}\to\Xi'^0_b\, e^+\nu_e$& 3.88
(1.66)&&$2.44\div3.28^\dagger$\\
$\Xi^{(2)\,+}_{cbu}\to\Xi'^0_b\, e^+\nu_e$&1.95 (3.91)\\
$\Xi^{(1)\,+}_{cbu}\to\Xi^{*\,0}_b\, e^+\nu_e$& 1.52 (3.45)\\
$\Xi^{(2)\,+}_{cbu}\to\Xi^{*\,0}_b\, e^+\nu_e$& 2.67 (1.02)\\
$\Xi^{(2)\,+}_{cbu}\to\Xi^0_b\, e^+\nu_e+\Xi'^0_b\, e^+\nu_e+\Xi^{*\,0}_b\, e^+\nu_e$& 7.27 (7.80)
&&&$(9.7\pm1.3)^*$\\
$\Xi^{*\,+}_{cbu}\to\Xi^{0}_b\, e^+\nu_e$&  4.08\\
$\Xi^{*\,+}_{cbu}\to\Xi'^{0}_b\, e^+\nu_e$&0.747\\
$\Xi^{*\,+}_{cbu}\to\Xi^{*\,0}_b\, e^+\nu_e$& 5.03\\\hline
\end{tabular}\hspace{.5cm}
\begin{tabular}{ll}
&{\hspace*{.375cm}$\Gamma \ [10^{-14}\,{\rm GeV}]$}\\
\hline
$\Omega^{(1)\,0}_{cbs}\to\Omega^-_b\, e^+\nu_e$&\hspace*{.5cm} 7.21 (3.12)\\
$\Omega^{(2)\,0}_{cbs}\to\Omega^-_b\, e^+\nu_e$&\hspace*{.5cm} 3.49 (7.12)\\
$\Omega^{(1)\,0}_{cbs}\to\Omega^{*\,-}_b\, e^+\nu_e$&\hspace*{.5cm} 2.98 (6.90)\\
$\Omega^{(2)\,0}_{cbs}\to\Omega^{*\,-}_b\, e^+\nu_e$&\hspace*{.5cm} 5.50 (2.07)\\
$\Omega^{*\,0}_{cbs}\to\Omega^-_b\, e^+\nu_e$&\hspace*{.5cm} 1.35\\
$\Omega^{*\,0}_{cbs}\to\Omega^{*\,-}_b\, e^+\nu_e$&\hspace*{.5cm} 10.2\\\hline
\end{tabular}
\end{center}
\end{table}
%
%\vspace{-1cm}
\begin{table}[h!!!]
\caption{Same as Table~\ref{tab:resctos} for  
$c\to d$ decays.  We have used
$|V_{cd}|=0.2252$. 
% Similar results are obtained
%for decays into
%$\mu^+\nu_\mu$.
}
\label{tab:resctod}
\begin{center}
\begin{tabular}{ll}
&{\hspace*{.5cm}$\Gamma \ [10^{-14}\,{\rm GeV}]$}\\
\hline
$\Xi^{(1)\,+}_{cbu}\to\Lambda^0_b\, e^+\nu_e$&\hspace*{.5cm} 0.219 (0.196)\\
$\Xi^{(2)\,+}_{cbu}\to\Lambda^0_b\, e^+\nu_e$&\hspace*{.5cm}  0.136 (0.154)\\
$\Xi^{(1)\,+}_{cbu}\to\Sigma^0_b\, e^+\nu_e$&\hspace*{.5cm}  0.198 (0.0814)\\
$\Xi^{(2)\,+}_{cbu}\to\Sigma^0_b\, e^+\nu_e$ &\hspace*{.5cm} 0.110 (0.217)\\
$\Xi^{(1)\,+}_{cbu}\to\Sigma^{*\,0}_b\, e^+\nu_e$&\hspace*{.5cm}  0.0807 (0.184)\\
$\Xi^{(2)\,+}_{cbu}\to\Sigma^{*\,0}_b\, e^+\nu_e$&\hspace*{.5cm}  0.147 (0.0556)\\
$\Xi^{*\,+}_{cbu}\to\Lambda^{0}_b\, e^+\nu_e$&\hspace*{.5cm}   0.235\\
$\Xi^{*\,+}_{cbu}\to\Sigma^{0}_b\, e^+\nu_e$&\hspace*{.5cm}  0.0399\\
$\Xi^{*\,+}_{cbu}\to\Sigma^{*\,0}_b\, e^+\nu_e$&\hspace*{.5cm}  0.246\\\hline
\end{tabular}\hspace{2cm}
\begin{tabular}{ll}
&{\hspace*{.5cm}$\Gamma \ [10^{-14}\,{\rm GeV}]$}\\
\hline
$\Omega^{(1)\,0}_{cbs}\to\Xi^-_b\, e^+\nu_e$&\hspace*{.5cm}  0.179 (0.164)\\
$\Omega^{(2)\,0}_{cbs}\to\Xi^-_b\, e^+\nu_e$&\hspace*{.5cm}  0.120 (0.133)\\
$\Omega^{(1)\,0}_{cbs}\to\Xi'^-_b\, e^+\nu_e$&\hspace*{.5cm}  0.169 (0.0702)\\
$\Omega^{(2)\,0}_{cbs}\to\Xi'^-_b\, e^+\nu_e$&\hspace*{.5cm}  0.0908 (0.182)\\
$\Omega^{(1)\,0}_{cbs}\to\Xi^{*\,-}_b\, e^+\nu_e$&\hspace*{.5cm}  0.0690 (0.160)\\
$\Omega^{(2)\,0}_{cbs}\to\Xi^{*\,-}_b\, e^+\nu_e$&\hspace*{.5cm}  0.130 (0.0487)\\
$\Omega^{*\,0}_{cbs}\to\Xi^-_b\, e^+\nu_e$&\hspace*{.5cm}  0.196\\
$\Omega^{*\,0}_{cbs}\to\Xi'^-_b\, e^+\nu_e$&\hspace*{.5cm}  0.0336\\
$\Omega^{*\,0}_{cbs}\to\Xi^{*\,-}_b\, e^+\nu_e$&\hspace*{.5cm}  0.223\\\hline
\end{tabular}
\end{center}
\end{table}
%\vspace{-20pt}
%
\section{Heavy Quark Spin Symmetry constraints on the form factors}
\begin{figure}
\centering
 \rotatebox{270}{ \resizebox{!}{6.5cm}{\includegraphics{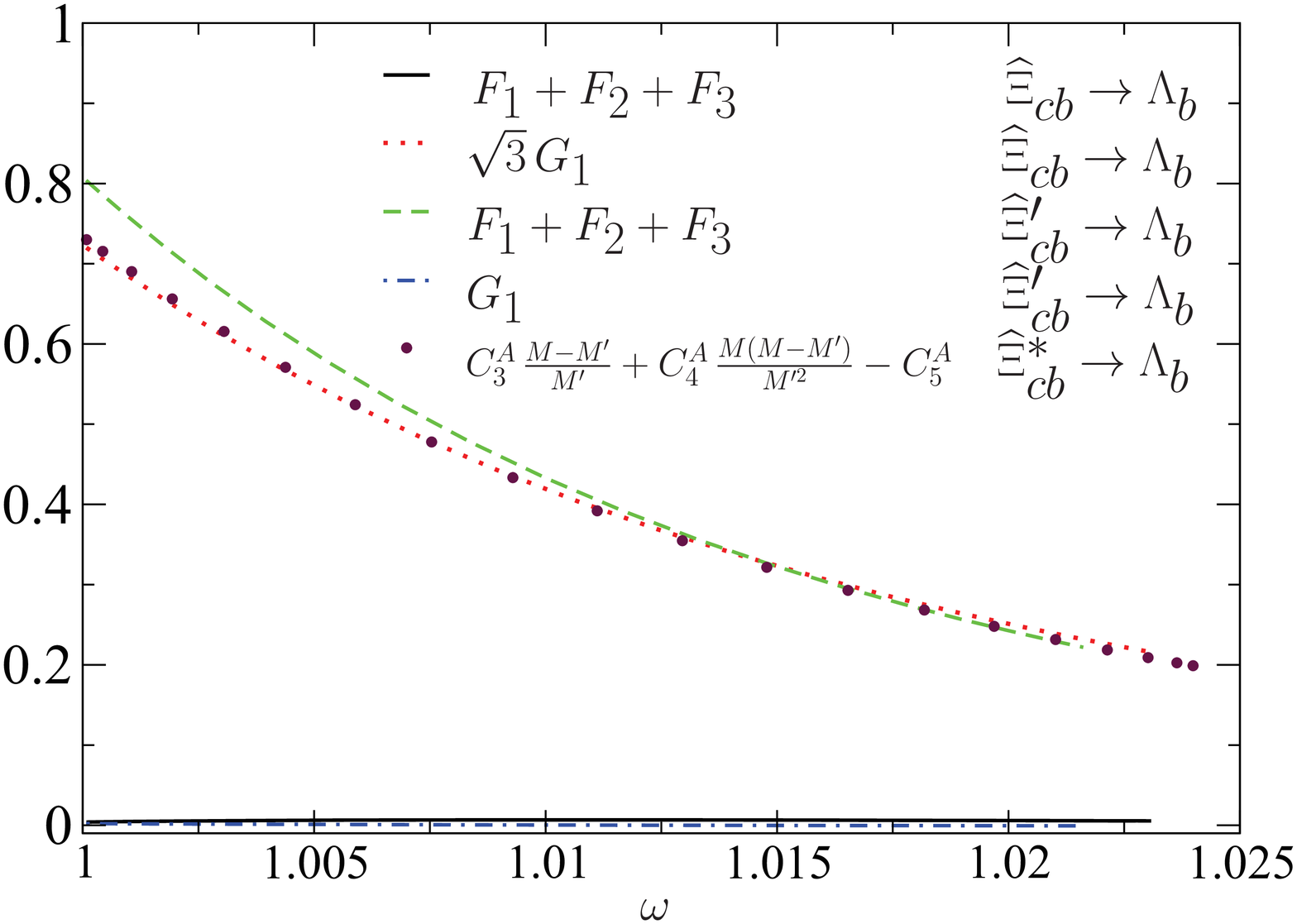}}}\hspace{0.5cm}
  \rotatebox{270}{\resizebox{!}{6.5cm}{\includegraphics{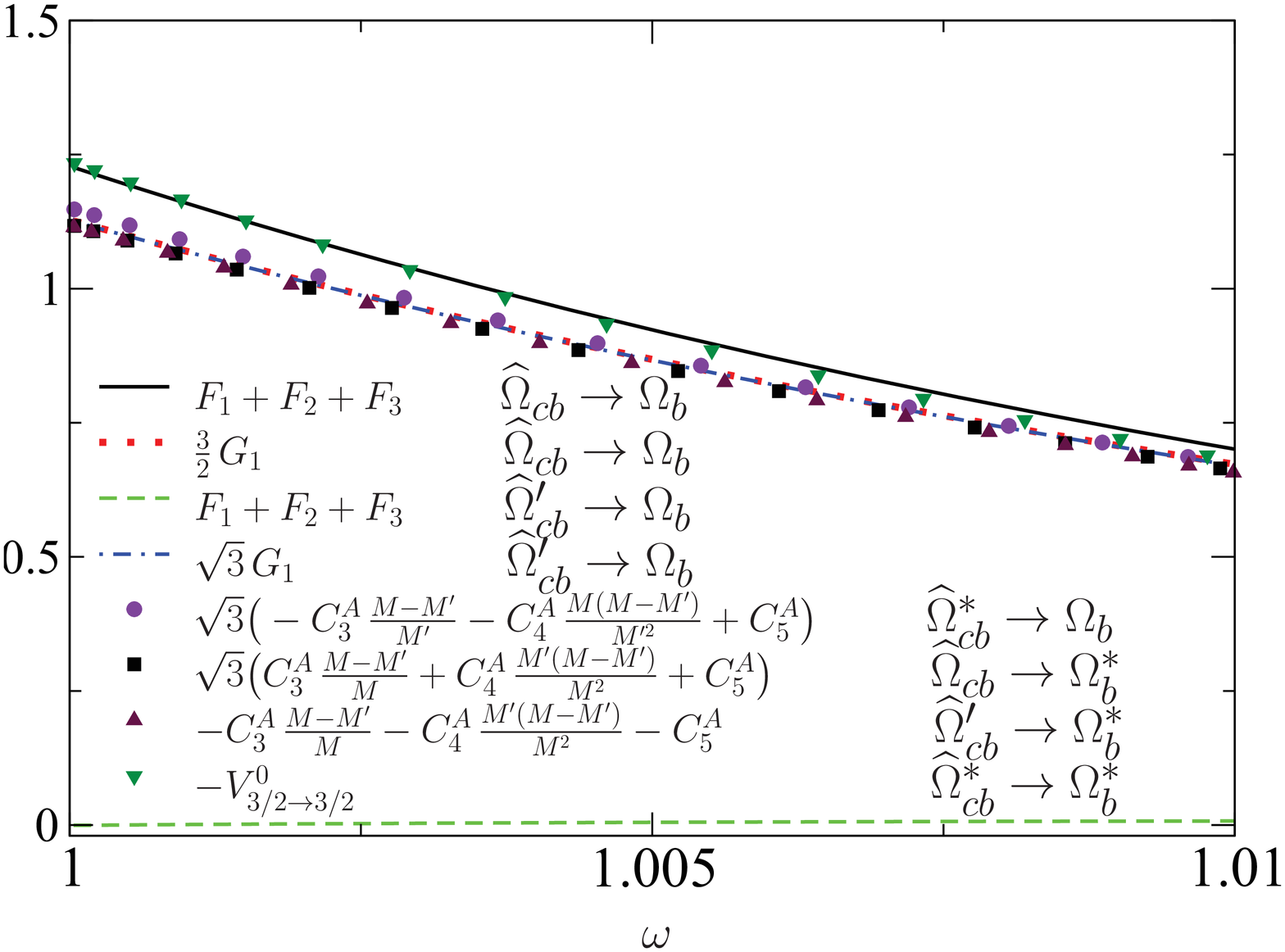}}}
\caption{Combinations of  form factors that 
are constrained by HQSS as explained in the text.}
\label{fig:1}       % Give a unique label
\end{figure}

Heavy quark spin symmetry (HQSS) imposes a number of constrains among
the form factors.  Although these constraints are approximate, they
can be used to make model independent predictions. These constraints
are consequences of the spin symmetry for infinite heavy quark masses
and can be derived using the trace formalism
~\cite{Falk:1990yz,Flynn:2007qt}. This is explained in detail in
Sec.~IV of Ref.~\cite{Albertus:2012jt}. As an example we shall mention
here some of the relations among form factors that hold at zero
recoil, i.e. at $\omega=1$, where $\omega$ is the product of the
initial and final meson four-velocities.  The constraints have been
obtained for transitions involving states in which $S_{cq}$ is well
defined. Note the $\hat B^*_{cb}$ is the same as $ B^*_{cb}$. For
$\hat B_{cb},\,\hat B'_{cb}$ or $\hat B^*_{cb}$ transistion to
$\Lambda_b$ or $\Xi_b$ baryons one obtains the relations
\begin{eqnarray}
&&\hat B_{cb}\to \Lambda_b,\Xi_b:\hspace{1cm} F_1+F_2+F_3=0, \hspace{1cm}
G_1=\frac{1}{\sqrt3}\eta\nonumber\\
&&\hat B'_{cb}\to \Lambda_b,\Xi_b:\hspace{1cm} F_1+F_2+F_3=\eta, \hspace{1cm}
G_1=0\nonumber\\
&&\hat B^*_{cb}\to \Lambda_b,\Xi_b:\hspace{1cm} -C_3^A\frac{M-M'}{M'}-C_4^A\frac{M(M-M')}{M'^2}+C_5^A=-\eta
\end{eqnarray}
where $\eta$ is the corresponding Isgur-Wise function. This function
is different for different ligth quark configurations in the initial
and final baryons.  Deviations from the above relations are expected
for finite heavy quark masses.  In the left panel of Fig.~\ref{fig:1}
we see our form factors approximately satisfy those constraints over
the whole $\omega$ range available for the transitions. Similarly, for
transitions from initial $\hat B_{cb},\,\hat B'_{cb},\,\hat B^*_{cb}$
to final $\Sigma_b^{(*)},\Xi^{\prime},\,\Xi^*,\,\Omega^{(*)}$ baryons
we have the zero recoil constraints
\begin{eqnarray*}
&&\hat B_{cb}\to \Sigma_b,\Xi'_b,\Omega_b:\hspace{1cm} F_1+F_2+F_3=\beta;
\hspace{1cm} G_1=\frac{2}{3}\beta\nonumber\\
&&\hat B'_{cb}\to \Sigma_b,\Xi_b,\Omega_b:\hspace{1cm} F_1+F_2+F_3=0;
\hspace{1cm} G_1=\frac1{\sqrt3}\beta
\end{eqnarray*}
\begin{eqnarray}
&&\hat B^*_{cb}\to \Sigma_b,\Xi'_b,\Omega_b:\hspace{1cm}
-C_3^A\frac{M-M'}{M'}-C_4^A\frac{M(M-M')}{M'^2}+C_5^A=\frac1{\sqrt3}\beta\nonumber\\
&&\hat B_{cb}\to \Sigma^*_b,\Xi^*_b,\Omega^*_b:\hspace{1cm}
C_3^A\frac{M-M'}{M'}C_4^A\frac{M(M-M')}{M'^2}+C_5^A=\frac1{\sqrt3}\beta\nonumber\\
&&\hat B'_{cb}\to \Sigma^*_b,\Xi^*_b,\Omega^*_b:\hspace{1cm}
C_3^A\frac{M-M'}{M'}C_4^A\frac{M(M-M')}{M'^2}+C_5^A=-\beta
\end{eqnarray}
For $\hat B*_{cb}\to \Sigma^*_b,\Xi^*_b,\Omega^*_b$ we get that the
matrix element of the zero component of the vector part of the weak
current equals $-\beta$ when evaluated between states with the same
spin projection. As before, $\beta$ is the corresponding Isgur-Wise
function which depends on the initial and final light quark
configurations.  Again, as shown in the right panel of
Fig.~\ref{fig:1}, we see the above HQSS constraints are approximately
satisfied over the whole $\omega$ range.
\begin{acknowledgements}
 This research was supported by  the Spanish Ministerio de Econom\'{\i}a y 
 Competitividad and European FEDER funds
under Contracts Nos. FPA2010-
21750-C02-02,  FIS2011-28853-C02-02,  and CSD2007-00042, by Generalitat
Valenciana under Contract No. PROMETEO/20090090, by Junta de Andalucia under 
contract FQM-225
and by the EU HadronPhysics3 project, Grant Agreement
No. 283286. 
\end{acknowledgements}

\bibliographystyle{unsrt}
\bibliography{biblist1.efb22}{}

\end{document}